\documentclass[fleqn,usenatbib]{mnras}

\usepackage{newtxtext,newtxmath}

\usepackage[T1]{fontenc}
\usepackage[normalem]{ulem}

\DeclareRobustCommand{\VAN}[3]{#2}
\let\VANthebibliography\thebibliography
\def\thebibliography{\DeclareRobustCommand{\VAN}[3]{##3}\VANthebibliography}

\usepackage{graphicx}	
\usepackage{amsmath}	
\usepackage{natbib}
\usepackage{physics}

\usepackage{color}
\newcommand{\eq}[1]{(\ref{#1})}
\newcommand{\Eq}[1]{equation~(\ref{#1})}
\newcommand{\Eqs}[2]{equations~(\ref{#1}) and~(\ref{#2})}

\newcommand{\Sec}[1]{\S\ref{#1}}

\newcommand{\Fig}[1]{Fig.~\ref{#1}}
\newcommand{\Tab}[1]{Table~\ref{#1}}

\newcommand{\mps}{m~s$^{-1}$}

\newcommand{\cmss}{cm$^2$~s$^{-1}$}

\def\bl{Babcock--Leighton}

\newcommand{\vect}[1]{\boldsymbol{#1}}

\title[Polar field and solar cycle variability]{Variabilities in the polar field and solar cycle due to  irregular properties of Bipolar Magnetic Regions}
\author[Kumar et al.]{
Pawan Kumar\thanks{E-mail: pawan.kumar.rs.phy18@itbhu.ac.in},  Bidya Binay Karak\thanks{E-mail: karak.phy@itbhu.ac.in}, Anu Sreedevi\thanks{E-mail: anubsreedevi.rs.phy20@itbhu.ac.in}
\\
Department of Physics, Indian Institute of Technology (Banaras Hindu University), Varanasi 221005, India
}

\date{Accepted XXX. Received YYY; in original form ZZZ}

\pubyear{2024}

\begin{document}
\label{firstpage}
\pagerange{\pageref{firstpage}--\pageref{lastpage}}
\maketitle

\begin{abstract}
Decay and dispersal of the tilted Bipolar Magnetic Regions (BMRs) on the solar surface are observed to produce the large-scale poloidal field, which acts as the seed for the toroidal field and, thus, the next sunspot cycle.
However, various properties of BMR, namely, the tilt, time delay between successive emergences, location, and flux, all have irregular variations.
Previous studies show that these variations can lead to
changes in the polar field.
In this study, we first demonstrate that our 3D kinematic dynamo model, STABLE, reproduces the robust feature of the surface flux transport (SFT) model, namely the variation of the generated dipole moment with the latitude of the BMR position. Using STABLE in both SFT and dynamo modes, we perform simulations by varying the individual properties of BMR and keeping their distributions the same in all the cycles as inspired by the observations. 
We find that randomness due to the distribution in either the time
delay or the BMR latitude produces negligible variation in the polar field and the solar cycle. However, randomness due
to BMR flux distribution produces substantial effects, while the scatter in the tilt around Joy’s law produces the largest
variation. Our comparative analyses suggest that the scatter of BMR tilt around Joy’s law is the major cause of variation
in the solar cycle.
Furthermore, our simulations also show that the magnetic field-dependent time delay of BMR emergence produces more realistic features of the magnetic cycle, consistent with observation.
\end{abstract}




\section{Introduction}
\label{Sec: int}
The 11-year magnetic cycle in the sun is not regular. 
The extreme variation in the cycle can be seen as grand minima such as the Maunder minimum and grand maxima namely the modern maximum \citep{Sol04, Sval13, Uso17, AB23}.
Besides this long-term modulation, there are short-term (6 -- 18 months) variations which are so-called bursts of activity or seasons of the sun \citep{Rieger, EG16} 
and the double peaks \citep{KMB18} in the solar cycle.
In addition to the cycle strength, the duration also 
varies in each cycle. 
\citet{wald1935} reported that the strong cycles take less time to rise and vice versa \citep[also see][]{KC11, CS16}. 
Hence, the variation in the amplitude and duration of the cycle is somewhat related.
Further, the time of reversal of the polar field of the sun is also not fixed, and it differs from cycle to cycle by a few months to years. In fact, there are evidences of possible triple reversals of the polar field \citep{MFS83, Mord22}. 
Because of this variable nature of the solar cycle, it is challenging to make a prediction of the future solar cycle \citep{Petrovay20, Bhowmik23}.
However, prediction is compelling because the solar magnetic cycle affects the heliosphere, thus our technology-dependent society \citep{Temmer21}.

 It is believed that the solar magnetic cycle 
 is the result of a dynamo operating in the solar convection zone (CZ) 
 and thus the variability observed in the solar cycle must be connected to the dynamo action \citep{Karak23}.
 There are enough evidences that the solar dynamo is of $\alpha$$\Omega$ type \citep{Kar14a, CS15, Cha20} in which 
 the $\alpha$ denotes the generation of poloidal field 
 from the toroidal one (the $\alpha$ effect) 
 and the ${\rm\Omega}$ represents the $\rm\Omega$ effect in which the toroidal field is generated from the poloidal one through the differential rotation.
Since the idea of \citet{Ba61} and \citet{Le64}, it has been observed that the tilted bipolar magnetic regions (BMRs) decay and disperse on the solar surface and produce large-scale poloidal field in the Sun \citep{Das10, KO11}. Hence, we shall consider this \bl\ process for the generation of the poloidal field in our solar dynamo model.
The poloidal field generated near the surface at low latitudes through the \bl\ process is advected towards the high latitudes and further to the deep  CZ with the help of meridional flow, where the $\Omega$ effect stretches the poloidal field to give rise to the toroidal field. This toroidal field due to magnetic buoyancy emerges on the solar surface in the form of BMRs \citep{Pa55b}.

In the solar dynamo model, the \bl\ process is a  crucial component 
which involves randomness. 
The extensive randomness arises in the 
\bl\ process through the tilt angle of BMR.
The tilt angle of BMR is observed to increase statistically with the increase of the latitude, which is known as Joy's law \citep{Hale19}. 
Observations also show that there is a considerable amount of scatter in tilt angle around this Joy's law \citep{How91, Jha20}. 
While in most of the previous studies, the tilt is measured from the sunspot group \citep[when magnetograms data are not available;][]{Sv07}  or BMR by repeatedly observing the same feature \citep[e.g.,][]{MN16}, recently \citet{sreedevi23} have tracked the BMRs for the last two solar cycles and produced unique tilt for each BMR during this period. 
Figure~2 of \citet{sreedevi24} shows Joy's law and the scatter around the tilt obtained from those tracked BMRs.
Observations show that the slope of Joy's law has some cycle-to-cycle variation, which can lead to some change in the polar field as demonstrated by \citet{BMN04}. However, the observed large scatter in the tilt around  Joy's law is another major cause of fluctuations in the poloidal field \citep{JCS14, KM17, LC17}.
Furthermore, observations show that about 8$\%$ of BMRs in a solar cycle are anti-Hale \citep{McClintock+Norton+Li14} 
while 25 to 30$\%$ are anti-Joy.
These anti-Hale and anti-Joy BMRs 
produce opposite polarity field and thus they disturb the regular polar field \citep{Ca13, JCS14, JCS15, KM18, Kit18, KMB18, Mord22, pal23}.  
\citet{Nagy17} showed that in the extreme case, a sufficiently complex wrongly tilted (rogue) BMRs can considerably disturb the dynamo and it even may lead to a grand minimum.

The other sources of irregularity in the solar dynamo are the time delay in the BMR emergence, latitudinal position, and the flux of BMR. It has been observed that the time delay between two successive BMRs is not the same, and it follows a distribution.
Again from the tracked BMRs from MDI and HMI for the last two solar cycles \citep{sreedevi23}, as presented in \Fig{fig:dist}(a), we find that the lag between two successive BMR emergences follow a distribution which can be approximated by a log-normal distribution.  This distribution in the time delay of BMR emergence can introduce a variation in the solar cycle.
A related time delay in the solar dynamo is the delay in the rise of the BMR-forming toroidal flux tube from the base of the CZ to the surface; see Sects. 5.3 and 6.4 of \citet{Karak23}. When this delay is magnetic field dependent, it can lead to a variation in the solar cycle \citep{JPL10}.
Furthermore, the latitude of BMR emergence on the solar surface is also not uniform and plays a role in determining the strength of the polar field \citep{BMN04, Nagy17}. In fact, \citet{JCS15} had shown that an anomalous BMR (anti-Hale or anti-Joy) appearing at low latitude produces a significant change in the dipole moment.
The mean latitudes of BMRs in a cycle depend on its strength; in the strong cycles, BMRs begin at high latitudes \citep{Wald1955, SWS08, JCS14, MKB17}.
 \citet{BMN04} showed the photospheric magnetic field at solar minimum decreases with the increase of the mean latitude of the BMR band. 
Finally, the flux content in the erupting BMR is not constant, and it randomly varies roughly in the range of $10^{21}$ to $10^{23}$ Mx following a log-normal distribution as seen in \Fig{fig:dist}(b); also see \citet{SJr66}.
Variation in the BMR flux can cause a change in the poloidal field.

Since the polar magnetic field or its proxy is strongly correlated with the activity level of the next cycle \citep{Sch78, CCJ07, WS09, KO11, Muno13, Priy14, Pawan21, Kumar22},
the variation of the polar field can cause a variation in the solar magnetic cycle. Therefore, to realize the behavior of the future solar cycle, it is essential to understand the 
polar field of the present cycle.

The variations that we are considering are essentially caused by the stochastic nature of the convection.
However, there is an extensive literature elucidating the long-term variability due to nonlinearities arising from the Lorentz feedback of the magnetic field on the flows \citep[e.g.,][]{Tob98, BTW98, WT16, Cha20, Karak23}.
However, nonlinearities do not always produce variation in the solar cycle. Some of the nonlinearities recognized in the \bl\ type 
dynamo models, such as the flux loss through magnetic buoyancy \citep{AB22}, latitude quenching \citep{J20, Karak20}, magnetic field dependent tilt \citep[so-called the tilt quenching][]{Das10, Jha20, sreedevi24}, inflows around active regions \citep{MC17, Nagy20, TM24} all have tendency of stabilizing the magnetic field rather than producing large variations in the solar cycle; see Sec. 6.1 of \citet{Karak23} for a detailed discussion.

In this work,  we shall only focus on the long-term variability in the solar cycle due to stochastic forcing caused by the convection on different parameters of \bl\ process, particularly the irregular properties of BMR.
We shall make comparative analyses of the variations of the polar field generated in a cycle arise by the randomness due to the distributions of
(i) the time delay of successive BMR eruptions, 
(ii) the BMR flux, 
(iii) the BMR tilt angle, and 
(iv) the latitude of BMR eruption.
Our studies differs from the previous ones who have also studied the effect of irregular properties of BMRs. For example, (i) \cite{BMN04} who studied the effect of the change of the polar field with the BMR flux distribution profile and shifting the mean latitude and width of the sunspot band, and the slope of Joy's law, 
(ii) \citet{Nagy17} who studied the effect of single rough BMR on the operation of the dynamo at different latitudes.
(iii) \citet{pal23} who showed the effect of anomalous (anti-Hale (anti-Joy), anti-Joy (Hale) and anti-Hale (Joy)) BMRs (5--10\% of the total flux) at different phases of the cycles with different latitudes.
 (iv) \citet{CJS16} and \citet{Bhowmik18} who studied the uncertainty in estimating the dipole moment and predicting the cycle strength by including the variability in BMR tilt, latitude  and flux.

\begin{figure}
\centering
\includegraphics[scale=0.4]{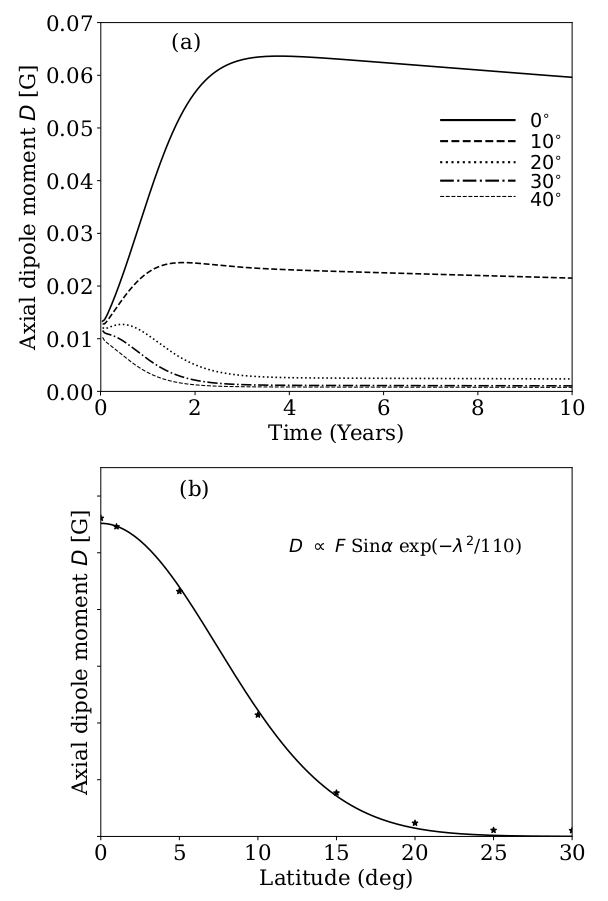}
\caption{The results produced from STABLE operating in SFT mode. (a) Time evolution of the axial dipole moment ($D$) of a BMR deposited at different latitudes with constant flux ($F$) of $10^{22}$ Mx and tilt $(\alpha)$ of $80^{\circ}$. (b) The values of the saturated (final) dipole moment vs the latitude of single BMR. 
The solid curve is the Gaussian profile of the form {\rm exp}($-\lambda^2$/110) as predicted by the SFT model of \citet{JCS14}.
}
\label{fig:dip}
\end{figure}

\begin{figure}
\centering
\includegraphics[scale=0.4]{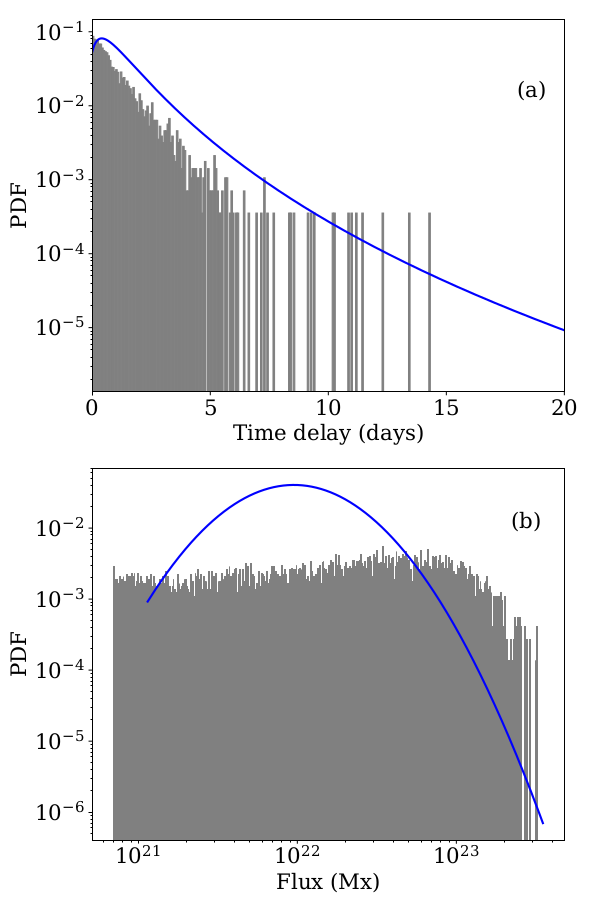}
\caption{Distributions of (a) the time delay between two successive BMR emergences and (b) flux content of 8800 BMRs obtained from MDI and HMI magnetograms during 1996--2020. These quantities are obtained by tracking each BMR throughout its lifetime or disk passage and taking the average values during the time when the flux is above 70\% of its maximum \citep{sreedevi23}. The solid lines are the log-normal distributions that approximately fit the data and are used in our (and  \citet{KM17}) calculations.
}
\label{fig:dist}
\end{figure}

Further, we shall perform simulations using synthetic BMR data of solar cycles deposited in our 3D kinematic dynamo model STABLE (Surface Flux Transport and \bl)
 and analyze the polar field and solar cycle variation.

\section{Model}
\label{sec:mod}
In the present work, we have used a 3D kinematic dynamo model STABLE \citep{MD14, MT16, HCM17, KM17}. In this model, BMRs are deposited based on the toroidal magnetic field present at the base of CZ, and the decay and dispersal of BMRs near the surface produce the poloidal field. The poloidal field eventually gives rise to the toroidal one through differential rotation and sustains the dynamo loop. This model can also be operated as a surface flux transport (SFT) model if we deposit BMRs on the solar surface. 
\citet{KM18} showed that this model also produces the correct latitude-dependent variation of the polar field---higher the latitude of BMR emergence, lower is the polar field generated. \citet{Karak20} showed that this effect leads to so-called latitude quenching which was suggested by \cite{Petrovay20, J20} based on the SFT model and observation.

To demonstrate the performance of the STABLE model as SFT explicitly, we present the axial dipole moment (DM) produced from a BMR deposited at different latitudes in \Fig{fig:dip}. We observe that the STABLE model also captures the behavior as seen in SFT model, namely, with increasing the latitude of the BMR emergence, the generation of DM decreases (\Fig{fig:dip}(a)) and the final saturated dipole moment produced can be approximated by a Gaussian profile as predicted by the SFT models \citep{JCS14, PNY20}.  We would like to mention that \citet{HCM17} failed to capture this feature as they did not include a downward pumping which is essential to make the model consistent with observations and SFT models \citep{Ca12, KC16}.   

\begin{figure}
\centering
\includegraphics[scale=0.4]{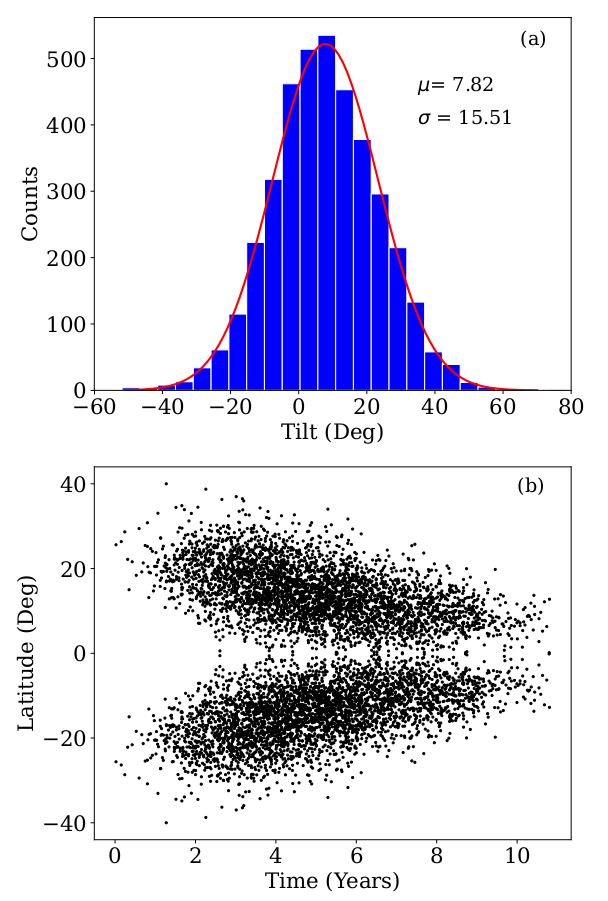}
\caption{(a) The distribution of BMR tilt with mean $= 7.82^{\circ}$ and scatter $\sigma = 15.51^{\circ}$ as consistent with observations \citep{wang15} and used in our study.  (b) Time-latitude distribution of synthetic BMR for one cycle generated using the algorithm described in \Sec{subsec:mod}.
}
\label{fig:dist1}
\end{figure}

In the present study, we shall exploit this feature of the STABLE dynamo model and utilize it as a dynamo simulator to study the solar cycle variation and as an SFT model to study the polar field evolution.
One advantage of using STABLE for our study is that we can use the same model in two modes, namely, SFT to study the surface behaviour of magnetic field and dynamo to study the global internal dynamics of fields.
In the next subsection, we first describe the STABLE model.  
\subsection{STABLE model}
In this model, we solve the Induction equation in three dimensions $(r,\theta,\phi)$ for the whole CZ with 
$0.69R \le r \le R$ 
($R$ is the radius of the sun), $0 \le \theta \le \pi$, and $0 \le \phi \le 2\pi$.
\begin{equation}
\frac{\partial \vect{B} }{\partial t} = \vect{\curl} \left[ (\vect{V} +\vect{\gamma})
\times \vect{B} - \eta_t \vect{\curl} \vect{B} \right],
\label{eq:ind}
\end{equation}
where $\vect{V}$ is the velocity field such that 
\begin{equation}
\vect{V} = v_r(r,\theta)\hat{r} + v_\theta(r,\theta)\hat{\theta} + r\sin\theta \Omega(r,\theta)\hat{\phi}.
\end{equation}
Here the axisymmetric velocity field is composed of meridional circulation $v_r, v_\theta$, and the differential rotation ($\Omega =  v_\phi / r\sin\theta)$. 
The profile of $\Omega$ used in this model 
roughly captures the observed properties as inferred through 
helioseismology \citep{Schou98} and was used in \citet{KM17}.
For the meridional circulation, we have considered a single cell circulation profile as used in \citet{KC16}.
In \Eq{eq:ind}, $\gamma$ represents the magnetic pumping which helps to suppress the loss of the toroidal field through the surface and thus boosts the dynamo efficiency even at high diffusivity \citep{Ca12, KC16}. 
Our model includes a radially downward magnetic pumping of speed 20~\mps\ in the near-surface layer $(r\ge 0.9R)$ of the sun.
In this model, we have considered a radial-dependent effective turbulent diffusivity $\eta_t$ of order $10^{12}$\cmss\ ($4.5 \times 10^{12}$\cmss for $r \ge 0.956R$ and $1.5 \times 10^{12}$\cmss\ below) throughout the whole CZ. Below the CZ, the magnitude of the diffusivity is reduced by about four orders of magnitude; 
see Eq. 4 of  \citet{KM17}.

A crucial part of this model is the Spotmaker algorithm, which places a BMR on the surface when certain conditions are satisfied.
The first condition for generating a BMR is that the magnetic field strength at the base of CZ must exceed a critical field strength $B_c$.
We have made $B_c$ latitude dependent such that it makes BMR emergence difficult at high latitudes and it helps the model BMRs to be consistent with the observations \citep{Karak20}.
The second condition is the time delay between successive BMR emergence.
After the first BMR eruption, the time delays for the subsequent eruptions are taken randomly from the observed log-normal distribution as shown in \Fig{fig:dist}(a), which follows the profile:
\begin{equation}
P(\Delta)= \frac{1} {\sigma_d\Delta \sqrt{2\pi}}{\rm exp}\left[\frac{-(\rm\ln\Delta - \mu_d)^2}{2\sigma_d^2}\right]
\label{eq:delay}
\end{equation}
Where 
$\sigma_d$ and $\mu_d$ are specified as,
$\sigma_d^2=\frac{2}{3}\left[\ln\tau_s - \ln\tau_p\right]$
and
$\mu_d=\sigma_d^2 + \ln\tau_p$.
Here $\tau_p=0.8$ and $\tau_s=1.9$ and
$\Delta$ is the time delay between two successive BMRs
(normalized to one day).

Now we have to specify the properties of BMRs (tilt, flux, and separation) in this model.
We take a log-normal distribution of flux that is close to the observed one as shown in \Fig{fig:dist}(b).
\begin{equation}
P(\phi)=\phi_0\frac{1}{\sigma_\phi\phi\sqrt{2\pi}}{\rm exp}\left[\frac{-(\rm\ln\phi -\mu_\phi)^2}{2\sigma_\phi^2}\right]
\label{eq:fl}
\end{equation}
where $\phi_0 = 1$, $\mu_\phi=51.2$, and $\sigma_\phi=0.77$.
We have specified the BMR field strength at the surface as 3 kG.
For the separation of BMRs, we choose the half distance between the centers of two spots of BMR to be 1.5 times the radius of the spot.
The radius of the BMRs is determined based on the 
log-normal distribution of the 
 flux and is given as,
$r=\sqrt{\frac{\phi}{B_s\pi}}$,    
where $B_s$ is the surface field.
For the tilt angle of BMRs, we follow the standard Joy's law \citep{Das10, SK12, sreedevi24}:

\begin{equation}
\delta = \delta_0 \cos\theta + \delta_{\rm f},
\label{eq:tilt}
\end{equation} where 
$\theta$ is co-latitude,  
$\delta_0=35^{\circ}$, and $ \delta_{\rm f}$ includes fluctuations in the tilt around Joy's law which follows a Gaussian distribution with $\sigma_\delta\approx15^{\circ}$ \citep[e.g.,][]{sreedevi24}. After giving a scatter of $\sigma_\delta\approx15^{\circ}$, we have found 30 -- 35\% BMRs are anti-Hale and anti-Joy. We have ensured that 25 -- $30\%$ 
of total BMRs would be anti-Joy, and about $8\%$ would be anti-Hale, as observational results suggest.
The distribution of the tilt $\delta$ for a cycle is shown in \Fig{fig:dist1}(a).

Further, we discuss the nonlinearity needed to limit the growth of the magnetic field. 
In our model, the latitude-dependent BMR threshold introduces a nonlinearity the so-called latitude quenching \citep{J20, Karak20}.
This helps to stabilize the growth of the magnetic field in our model.
In addition to this, we include a magnetic field-dependent quenching of the form:
1/$\left[1 + (B/B_0)^2\right]$ 
in the tilt of BMR (tilt-quenching), inspired by the observations \citep{Jha20}; 
for more details about the model see \citet{KM17}.

\subsection{Deposition of synthetic BMRs in STABLE}
\label{subsec:mod}
Now we discuss how instead of using the SpotMaker algorithm, we can feed synthetic BMR data into STABLE to operate this as SFT model.  
Following \citet{Jiang_2018}, we develop a synthetic BMR generation code (SBMR Code) to produce smoothed monthly synthetic BMR data. This can be described by the following equation.
\begin{equation}
\overline F_{sm}(t) = f_2(t) + \overline{\Delta r(t)}f_2(t).
\end{equation}
Here
\begin{equation}
f_2(t) = \frac{at^3}{\exp\left[\frac{t^2}{b^2}\right]-c},
\end{equation}
where $t$ is time in month, 
$a=\left(\frac{S_n}{9072.8}\right)^\frac{1}{0.706}$ with $S_n$ the amplitude of the sunspot cycle,
$b = 27.12 + \left(\frac{25.15}{1000 a}\right)^\frac{1}{4}$,
and $c = 0.71$.\\
When $t < 72$ months,
\begin{equation}
\overline{\Delta r(t)}=8.68 \exp\left(\frac{-z_1^2}{2}\right) 
\end{equation}
with $z_1 = \frac{t-200.44}{45.86}$.\\
When $72 \le t \le 108$ months, then
\begin{equation}
\overline{\Delta r(t)}=2.64 \exp\left(\frac{-z_2^2}{2}\right)
\end{equation}
with $z_2=\frac{t- 149.0}{27.31}$.\\ 
Finally in the last phase, when $t > 108$ months, 
the synthetic BMR follows the given profile:
\begin{equation}
\overline F_{sm}(t) = f_2(t).
\end{equation}

After getting synthetic BMRs, we need to specify the properties of these BMRs.
The mean latitude of the BMRs distribution $\lambda_n$ depends on the cycle strength $S_n$ and it   
follows the profile:
\begin{equation}
\lambda_n(t)=\left[26.4-34.2\frac{t}{T_{cy}}+16.1 \left(\frac{t}{T_{cy}}\right)^2\right]\frac{\overline\lambda_n}{14.6},
\label{eq:bflylat}
\end{equation}
where $\overline\lambda_n=12.2+0.015S_n$ and $T_{cy}$ is the cycle period in month.
The scatter in latitude distribution obeys a Gaussian profile with  
$\sigma$ 
defined as:
\begin{equation}
\sigma=\left[0.14+1.05\frac{t}{T_{cy}}-0.78\left(\frac{t}{T_{cy}}\right)^2\right]\lambda_n(t).
\label{eq:bflysigma}
\end{equation}
We consider synthetic BMR emergence symmetric in both hemispheres, and the longitudinal distribution of BMRs are random.
The BMR eruption time delay is taken randomly within a month. 
The time-latitude distribution of the resultant synthetic BMR data for one cycle is shown in \Fig{fig:dist1}(b).
The flux and tilt are taken from the same prescriptions as discussed above (\Eqs{eq:fl}{eq:tilt}). 

\section{Results and Discussion}
We first quantify the effect of randomness due to distributions in BMR properties on the polar field using the STABLE dynamo model as SFT. Later we shall 
measure the variability of the solar cycle using the STABLE as a dynamo model. 

\subsection{Variation in the polar field}
\begin{figure}
\centering
\includegraphics[scale=0.38]{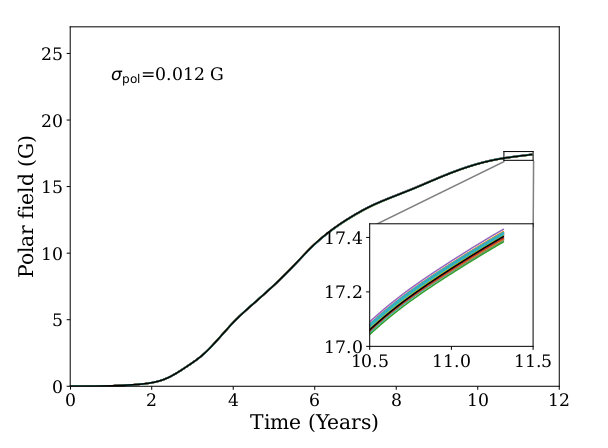}
\caption{The plot shows the variation in the polar field arises from the randomness due to the distribution in the time delay. The black curve is showing a reference polar field in the plot (see inset for the zoomed-in view). 
}
\label{fig:tim_fl}
\end{figure}

\begin{figure*}
\centering
\includegraphics[scale=0.4]{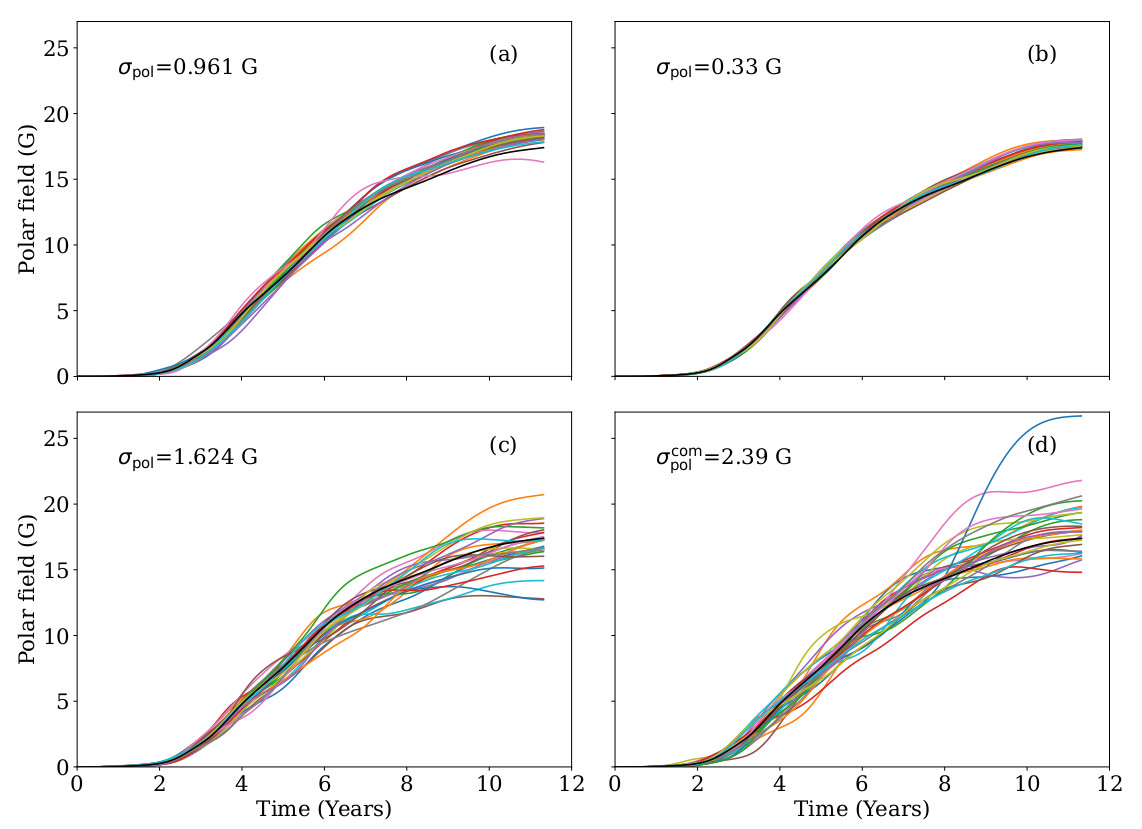}
\caption{Format of each panel is the same as \Fig{fig:tim_fl}, but shows the evolution of polar field from simulations with distribution in (a) BMRs flux (with $\sigma$ of distribution 0.77), (b) BMRs latitude (with $\sigma$ of the distribution given by \Eq{eq:bflysigma}), (c) BMR tilt (with distribution $\sigma = 15.51^{\circ})$, and (d) all four BMR parameters.
}
\label{fig:sub}
\end{figure*}

We feed synthetic BMR data for one cycle (of about 11~years long) in STABLE and perform simulations by operating it as SFT model with random variation in different BMR properties. 
For the statistical reliability of the result, we have 
repeated each case for 30 different realizations of the random number.
As a reference case, we first perform a simulation without adding any random component in the BMR properties. 
To address the variability in the polar field arising from different fluctuating parameters of BMRs, we calculate the standard deviation $\sigma_{\rm pol}$ from $30$ realizations in each case relative to the reference polar field. 
We also calculate the relative variability attributed from the distributions of various BMR parameters, denoted as 
$\sigma_{\rm pol} / \sigma_{\rm pol}^{\rm com}$, where $\sigma_{\rm pol}$ 
represents the standard deviation due to individual BMR parameters (e.g., for time delay
$\sigma_{\rm pol} = 0.012~G$), 
and
$\sigma_{\rm pol}^{\rm com}$
represents the standard deviation resulting from the combined fluctuations 
which is $2.39~G$); 
see \Tab{table1}.
For the quantification of the polar field, we compute the remnant magnetic flux strength from $55^{\circ}$ latitude to the pole in the northern hemisphere.
It is worth mentioning that our focus is solely on assessing 
the comparative effects of various irregular BMR parameters on the polar field.
Consequently, we make two hemispheres 
symmetric. However, \citet{Nagy17} studied the hemispheric asymmetry due to a single rough BMR and 
in our dynamo simulations, we do not impose hemispheric asymmetry.
The variation of this polar field for the reference case is shown by the black curve in all the Figures~\ref{fig:tim_fl} -- \ref{fig:sub}.

\begin{table}
\centering
\caption{Mean $\mu$ and standard deviation $\sigma$ of the distribution of various BMR parameters (2nd column), variability in the polar field $\sigma_{\rm pol}$ (3rd column) and the relative possible percentage variability (4th column) produced in the polar field  due to different individual fluctuation parameters of the BMRs relative to the combined fluctuations (\Fig{fig:sub}(d); 
$\sigma_{\rm pol}^{\rm com} = 2.39~G$).
}
\begin{tabular}{llclcl} 
\cline{1-4}
Variable & Distribution & $\sigma_{\rm pol}$ & Relative $\%$ \\
parameters & ($\mu ,\sigma$) & $(G)$ & variability \\

\cline{1-4}
Time delay &  0.35, 0.76 & 0.012 & 0.5\%   \\
    
\cline{1-4}

Latitude &  \Eqs{eq:bflylat}{eq:bflysigma} & 0.330 & 13.8\%    \\
\cline{1-4}
Flux  &  51.2, 0.77   & 0.961 &  40.2\%  \\
\cline{1-4}
Tilt scatter	&  $7.82^{\circ}$, $15.51^{\circ}$  & 1.624 & 68.7\%  \\
\cline{1-4}

\end{tabular}
\label{table1}
\end{table}

\subsubsection{Due to the distribution in BMR time delay} 
\label{sec:vari_tdelay}
The time evolutions of the polar fields obtained from the simulations with the different realizations of the time delay are shown by different colors in \Fig{fig:tim_fl}. 
We find that these curves deviate only negligibly from the reference polar field (see black curve and the inset) and the standard deviation from the reference polar field is $\sigma_{\rm pol} = 0.012~G$; see \Tab{table1}.
The reason for the tiny changes in the polar field is the following. The most probable time delay of the BMR emergence rate is less than a day (mode of the distribution is 0.8 days and mean is 1.9 days; see \Fig{fig:dist}(a) and the probability decreases with increasing the time delay. When the time delay is 10 days, the probability of the BMR emergence is less than 1\%. 
In contrast, the polar field takes about 1 to 2 years to reach the pole from the active latitude belt where it is formed. Hence, the variation in the polar field caused by the variation in the time delay is smoothed out by the time it reaches the pole and we observe only a negligible variation. 
We note that when we performed different simulations at different realizations of random delay, we kept the total number of BMRs the same in one month. Instead of this, if we increase this time to six months, then we also find a tiny variation in the polar field with respect to the reference case.  
Hence, based on these simulations, we conclude that the randomness in the time delay introduces only a slight variation in the polar field.

\subsubsection{Due to the distribution in BMR flux}
\label{sec:polvari_flux}
Now, we analyze the result of the simulations with variation in the BMR flux, which is lognormally distributed as mentioned in equation \eq{eq:fl} of \Sec{sec:mod}, keeping all other parameters of the BMR invariant. For 30 realizations, the total flux of sunspots in a solar cycle is considered a fixed value of $8.93\times 10^{25} Mx$.
\Fig{fig:sub} (a) shows the variation of the polar field for this case.
Now we observe a noticeable variation in the evolution of the polar field.
The standard deviation from the reference polar field is found to be $\sigma_{\rm pol} = 0.961~G$.
A variation, in this case, is expected because the contribution to the polar field in a given hemisphere from decayed active region flux would be proportional to the total hemispheric active region flux \citep{KO11, Petrie15}.
Moreover, \citet{SD12} studied that the net polar flux is also affected by the number and size of the large flux concentrations, which qualitatively agrees with our findings.
Therefore, the variation in the BMR flux is one cause of variation in the polar field.

\subsubsection{Due to the distribution in the BMR latitude}
\label{sec:latvary}
Variation in the latitude distribution of BMR also gives a variation in the polar field strength ($\sigma_{\rm pol} = 0.33~G$). This is shown in \Fig{fig:sub}(b), in which we present the time evolution of the polar field from simulations at different realizations of the BMR latitudes following the same distribution.
We note that the separation of the leading and trailing poles of the BMRs remains unchanged; only the mean (of leading and trailing) latitude of the BMR changes randomly in each cycle. While changing the mean latitude, the tilt is also changed according to Joy's law (\Eq{eq:tilt}); however, the fluctuations in tilt as parameterized by $\delta_f$ remain the same in all simulations. Thus arguably, the variation that we observe in \Fig{fig:sub}(b) is the combined effect of latitude and tilt.   
We further note that we keep the mean and spread of the latitude distribution of BMRs the same for all cycles and vary only the individual position of BMR randomly (\Eqs{eq:bflylat}{eq:bflysigma}). Thus, our investigation differs from \citet{BMN04} who studied the impact of BMR emergence latitude by changing the mean and spread of the latitude band. 
Despite the methodological differences, our results qualitatively align with \citet{BMN04} findings, supporting the conclusion that changes in the latitude of active regions influence the polar field's strength.
A variation in the polar field due to a change in the BMR latitude is expected because when BMRs appear at low latitudes, the cross-equatorial cancellation is much more effective than the BMRs which appear at higher latitudes \citep{DTW04, JCS15, pal23}, which has been demonstrated in \Fig{fig:dip}.  
Moreover, \citet{JCS14, KM18}
showed that the latitudinal variation of BMRs affects the polar field strength \citep[also see][]{Mord22}. We note that in our case, the variation in the polar field is less because the mean and the width of the latitude distribution remain the same in all the simulations, while in the observed solar cycle, these vary with the cycle strength. Thus, our study shows that even when we keep the mean latitudinal distribution of the BMR the same and vary the individual latitude of the BMR, we find a noticeable variation in the polar field.

\begin{figure*}
\centering
\includegraphics[scale=0.36]{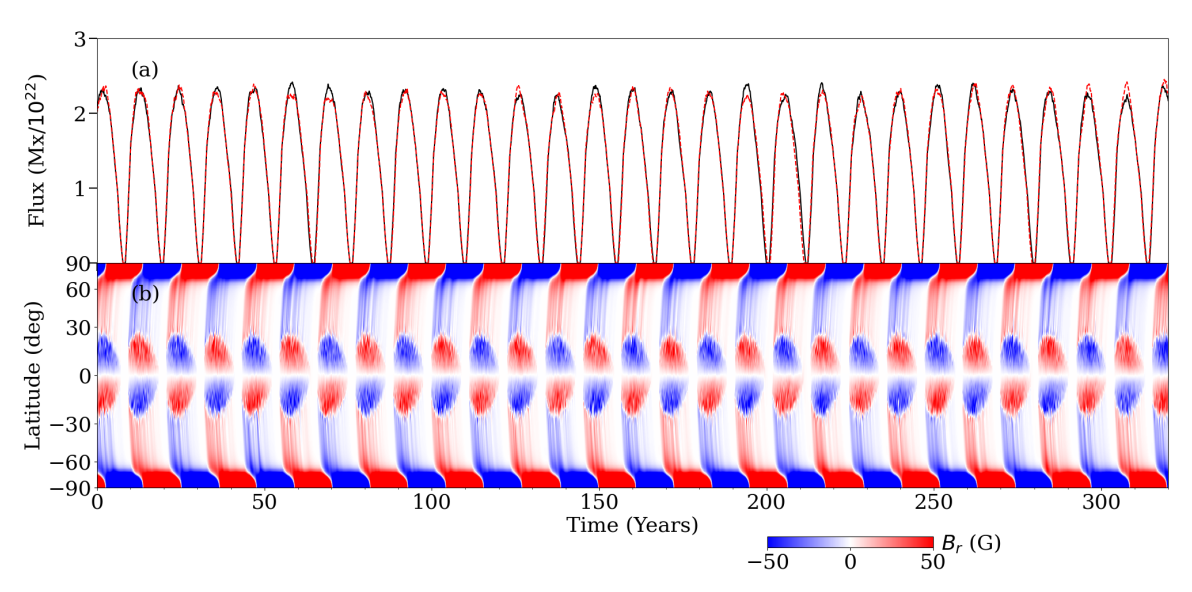}
\caption{Temporal variations of (a) the monthly total flux of BMRs (red/black: north/south) and (b) the azimuthal-averaged surface radial field 
from the simulations in which the time delay in BMR eruption is randomly taken from a distribution.
}
\label{fig:bflyC}
\end{figure*}

\subsubsection{Due to the distribution in BMR tilt around Joy’s law}
Now, we will demonstrate the effect of the tilt scatter of BMR on the polar field. We computed the tilt of BMR using \Eq{eq:tilt} i.e., Joy's law with Gaussian scatter captured by $\delta_f$.
In each simulation, $\delta_f$ changes randomly while keeping the distribution same. 
\Fig{fig:sub} (c) shows the effect of tilt scatter on the polar field variation ($\sigma_{\rm pol} = 1.624~G$). 
As in the \bl\ process, it is the tilt which makes the generation of the poloidal field possible in the Sun; any change in the tilt can certainly produce variation in the poloidal field. Particularly when we have a wrongly tilted BMR (anti-Joy and anti-Hale), it produces an opposite polarity field \cite[e.g.,][]{Nagy17, Mord22}.  
Recently, \citet{pal23} also showed that the emergence timing, relative numbers, latitude distribution, and flux content of anomalous active regions significantly impact the reversal timings and strength of the dipole moment; also see \citep{GA23, ABP23}. 
Furthermore, \citet{JCS15} also mentioned that the cause of the weaker cycle 24 is the wrong-tilted BMR (anti-Hale and anti-Joy) at the lower latitude, which decreases the polar field strength.
In our model, we have included about 30 to 35\%\ anti-Hale and anti-Joy BMRs (consistent with observations). 
These anti-Hale and anti-Joy BMRs have opposite tilts than regular BMRs and generate the opposite polarity field.  This opposite polarity field brings about a large fluctuation in the polar field of our simulation.
Previously \citet{Lisa21} also found that increasing the tilt scatter around Joy's law significantly increases the change in the polar axial dipole from one cycle to the next.
By comparing the simulations presented in the previous three sections with randomness in BMR time delay, flux, latitude, and tilt, we conclude that the tilt scatter produces maximum variation in the polar field. For comparative variability in the polar field due to these parameters (see \Tab{table1}).

\begin{figure*}
\centering
\includegraphics[scale=0.36]{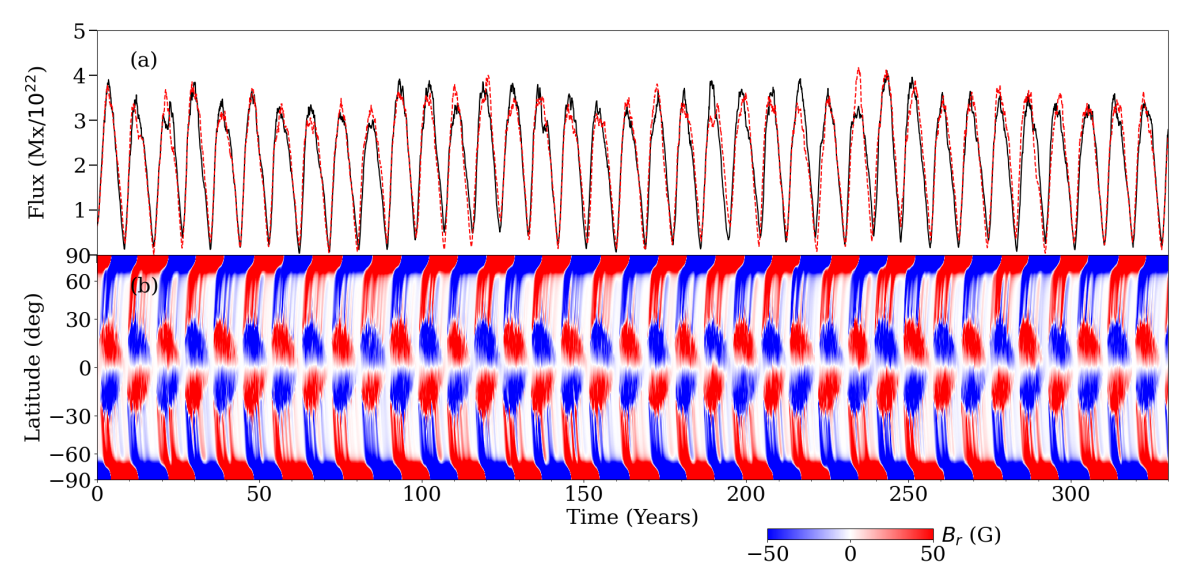}
\caption{Same as \Fig{fig:bflyC} but from the simulation in which BMR flux is taken from a distribution.}
\label{fig:bflyB}
\end{figure*}

\begin{figure*}
\centering
\includegraphics[scale=0.36]{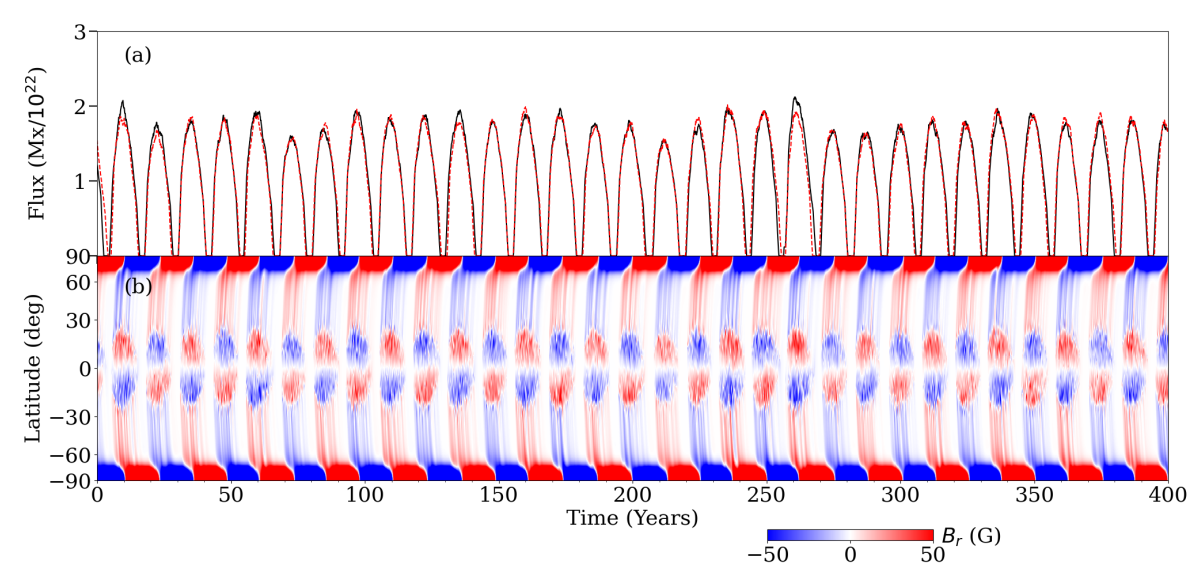}
\caption{Same as \Fig{fig:bflyC} but from the simulation in which the tilt is randomly taken from a distribution.}
\label{fig:bflyA}
\end{figure*}
\subsubsection{Due to combined fluctuations}
Now we measure the polar field variation due to the combined effect of all the fluctuations, i.e., the variations in tilt, time delay, flux, and latitude. This replicates the realistic scenario. 
The fluctuations due to the distributions of time delay, flux, latitude and tilt are captured precisely in the same proportion (distribution) as they were considered in the individual cases performed above.
\Fig{fig:sub} (d) shows the polar field variation in this case.
We find that in this case, the variation in the net polar field strength is highest 
($\sigma_{\rm pol} = 2.39~G$, which we define as $\sigma_{\rm pol}^{\rm com}$).
Maximum variation is expected because all individual random effects can contribute additively. 

\begin{figure*}
\centering
\includegraphics[scale=0.36]{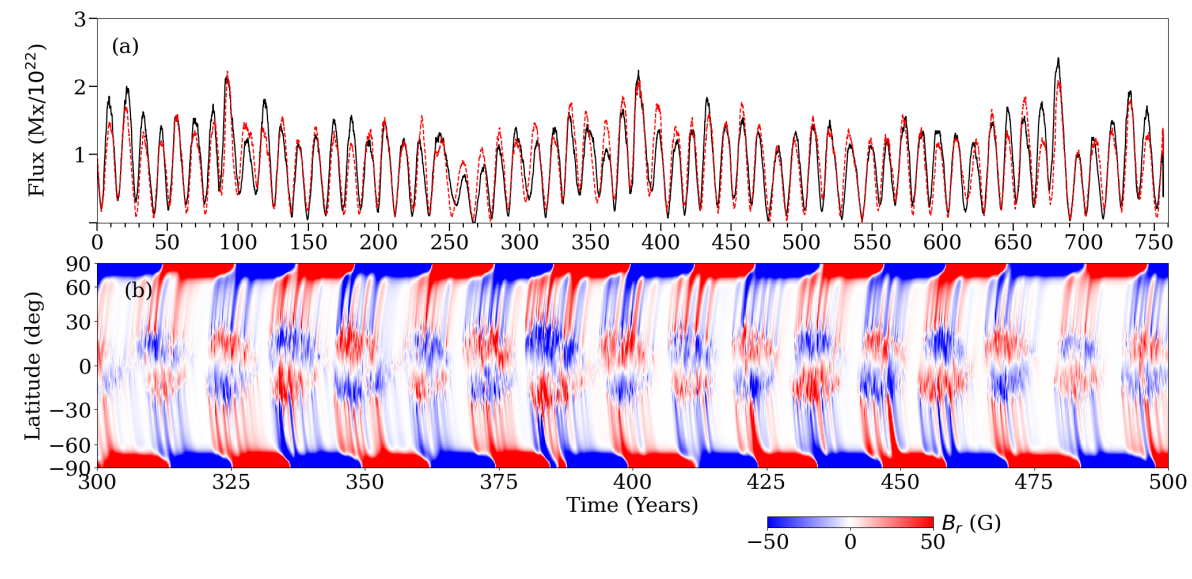}
\caption{Same as \Fig{fig:bflyC} but from simulations including distributions in all BMR properties (i.e., time delay, flux, latitude, and tilt) and the time delays are magnetic field independent.}
\label{fig:wmag}
\end{figure*}

\subsection{Variation in the solar cycle}
Now we study the variability of the solar cycle due to irregular properties of BMR by operating STABLE in default mode, i.e., as a dynamo model. For this, we consider three cases: 
(i) irregular time delay in the BMR eruption,  (ii) irregular BMR flux, and (iii) scatter in BMR tilt angle around Joy's law.
The incorporation of fluctuations in different BMR properties in the dynamo model is exactly the same as done in the study of polar field variation, which are explained in \Sec{sec:mod}. 
To be more specific, the time delay, flux and tilt distributions are taken from equations \eq{eq:delay}, \eq{eq:fl}, and \eq{eq:tilt}, respectively.
We note that to study solar cycle variation, we are not considering the case of the variation in the BMR latitude separately because it is not trivial to vary BMR latitude randomly by hand in the dynamo mode of this model. 
When the cycle strength varies, the mean latitude of BMR varies; hence it will be captured when cycle strength varies due to variation in the polar field caused by fluctuations in BMR properties.
In STABLE, at every numerical timestep, a BMR latitude is chosen randomly out of the latitude points where the azimuthal field exceeds the threshold which is latitude-dependent. 
Moreover, we have seen in \Sec{sec:latvary} (\Fig{fig:sub}(b)) that in the case of variation in latitude due to its distribution, the variation in the polar field is less compared to the variations in the tilt and flux and thus we are not performing any simulations for this case.

\Fig{fig:bflyC} shows the solar cycle variability produced by the random time delay in BMR emergence. The variation is measured in terms of the monthly value of the BMR flux and the surface radial field as the function of time.
We find that the variability in the solar cycle in both hemispheres is negligible.
In the polar field evolution \Sec{sec:vari_tdelay}, we have already seen that 
the random time delay in BMR eruption causes a negligible variation in the polar field. 
In the context of \bl\ dynamo, the poloidal field gives toroidal field, which eventually produces BMRs on the solar surface.
Therefore, a tiny variation in the polar field is expected to cause a  tiny variation in the next solar cycle. While this expectation is based on the dynamo mechanism \citep[e.g.,][]{CB11, Pawan21} and the observations \citep{Muno13}, if the dynamo operates in the chaotic region (much above the critical transition), even a small change in the polar field can lead to a large change in the solar cycle. However, it is important to note that our model does not operate within that specific region, and thus, such effects are not accounted for in our model.
In summary, the simulation performed in this case indicates that irregular rate of BMR emergence does not produce significant variation in the solar cycle.

\begin{figure*}
\centering
\includegraphics[scale=0.36]{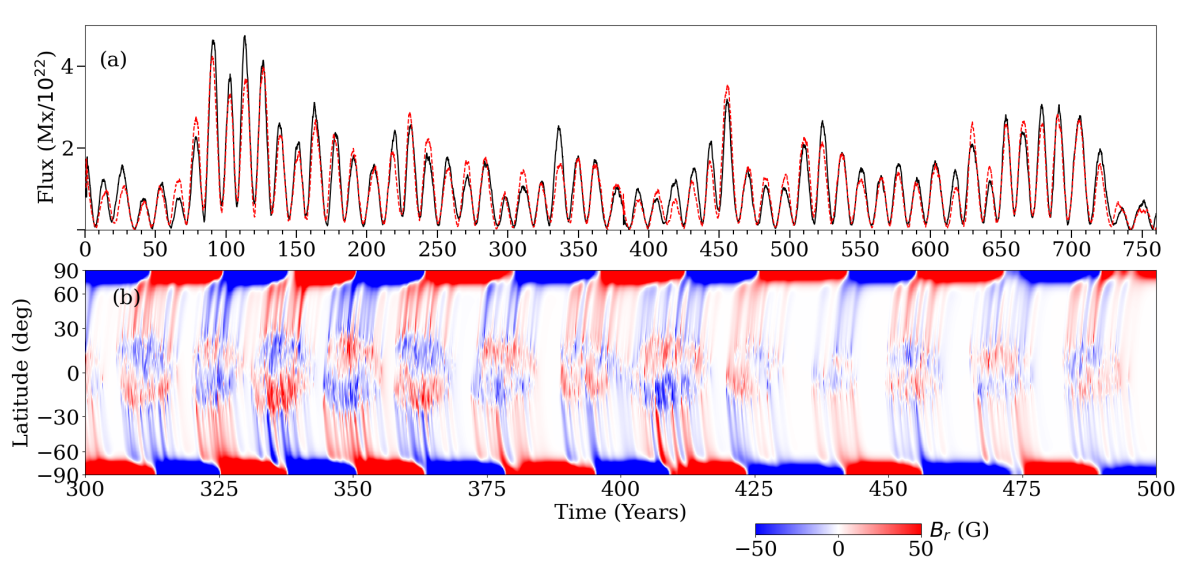}
\caption{Same as \Fig{fig:wmag} but with magnetic field-dependent time delay.}
\label{fig:mag}
\end{figure*}

However, when we perform the dynamo simulation with the variation in BMR flux, we observe a considerable amount of long-term modulation in the solar cycle as seen in \Fig{fig:bflyB}. 
This is expected given the fact that it produces a significant variation already in the polar field as seen in \Sec{sec:polvari_flux}.
Next, when we consider the scatter in BMR tilt around Joy's law, we again find  
a large amount of variation in the solar cycle as seen in \Fig{fig:bflyA}.
Again this result is congruous with the expectation that a considerable variation in the polar field is observed with the tilt scatter. 
Previous studies have also shown that the scatter in BMR tilt can produce large variations in the solar cycle, including grand minima and grand maxima \citep{LC17, Nagy17, KM17, KM18}.

Now we perform two other simulations 
by including all the fluctuations mentioned above individually i.e., variation in time delay, flux, and tilt. 
The basic difference between these two simulations is the only time delay. In one simulation, the time delay is magnetic field independent (\Fig{fig:wmag}), while in another it is magnetic field dependent (\Fig{fig:mag}).
To make the time delay magnetic field dependent, we make the following changes in the time delay parameters for both hemispheres:
\begin{equation}
\tau_p = \frac{2.2} {1 + \left[\frac{B_b}{400}\right]^2}, ~~~~{\rm{and}}~~~~
\tau_s = \frac{20} {1 + \left[\frac{B_b}{400}\right]^2},
\end{equation}
where $B_b$ is the azimuthal-averaged toroidal magnetic field in a thin layer spanning from $r = 0.715R$ to $0.73R$ at approximately $15^{\circ}$ latitude.

We find more variability and grand minima-like appearance when the time delay is magnetic field dependent. However, in the magnetic field independent case, we observe a few triple reversals of the polar field; see reversal in the southern hemisphere between 375 -- 400 years in  \Fig{fig:wmag}(b). 
We find triple reversal events in the magnetic field-independent time delay case because in this case, the emergence rate of BMRs does not change much over the solar cycle, while in the magnetic field-dependent case, it varies largely with the solar cycle---becoming highest around the solar maxima and lowest around the solar minima (because in this case the time delay is magnetic field dependent and varies according to the magnetic field strength). 
Therefore, in the magnetic field-independent case, the emergence rate around the cycle maxima is less (compared to the  magnetic field-dependent case), and as the BMR number is less, the effect of tilt scatter 
(the anti-Joy and anti-Hale BMR)
is statistically more prominent.
However, in the magnetic field-dependent case, the effect of tilt scatter largely cancels out due to the relatively large number of BMR emergences at the times of cycle maxima and this does not lead to triple reversals.
Moreover, observations and models \citep{Mord22} both reveal that the occurrence probability of triple reversal events is maximum around the time of solar cycle maxima.

Further, magnetic field-dependent and independent time delays produce variability in the solar cycle in a different way. 
Since grand minima and maxima-like events are more probable in the magnetic field-dependent time delay case and the morphology of the magnetic field better resembles the solar observation in this case, the magnetic field-dependent time delay case is more favorable with observation. Indeed, the distribution of the time delay in the observation varies with the solar cycle---becoming narrower during solar maximum and thus it is magnetic-field dependent.
\section{Conclusions}
In this study, we have utilized the 3D STABLE dynamo model to quantify the relative variations of the polar field and the solar cycle due to the stochastic properties of BMRs in the poloidal field generation (\bl\ process). 
We, for the first time, showed that our dynamo model can reproduce the most robust feature, namely the variation of the generated axial dipole moment with the latitude of BMR, as found in SFT models \citep[e.g.,][]{JCS14}. 
Thus, it gives some confidence on the applicability of the results of the dynamo model to observations. 
Next we present the distributions of the flux and time delay of BMRs obtained by tracking them from the high-resolution magnetograms.
Guided by these observations that the BMR properties follow distributions, we have 
studied 
the variations that arise from the distributions in the following processes:  (i) time delay in BMR emergence, (ii) BMR flux, (iii) BMR latitude, and (iv) BMR tilt angle. While earlier studies presented significant disturbance in the polar field due to the scatter in the BMR tilt around Joy's law \citep[e.g.,][]{JCS14, KM17}, anomalous BMRs \citep{Nagy17, pal23}, changes in the flux and latitudinal positions of BMRs \citep{BMN04, HCM17, Bhowmik18, pal23}, here we presented a comparative analysis of the variation of the polar field and the solar cycle strength using the same input and the same model (STABLE), operating as SFT and dynamo modes. 
Unlike previous studies, we have kept the cycle strength same, and thus the mean and width of the distributions of the tilts, latitudes, and fluxes of BMRs of the cycles are the same, and we only varied the individual properties of BMRs, keeping their distributions the same. 
The most surprising result of our study is that 
the time delay and latitude variations produce only little variations in the mean polar field of the cycle (and thus the cycle strength). A relatively large variation is observed due to distribution in the flux and the largest variability appears due to scatter in the BMR tilt; see \Tab{table1}. 
The combined effects of 
the distributions of all parameters of the BMRs on the polar field and the solar cycle are obviously the highest, and they are sufficient to reproduce a large spectrum of the observed modulation in the solar cycle and the magnetic field, including producing 
grand minima and maxima, and triple reversals of the polar field \citep{Mord22}.
In agreement with previous results, our study suggests that the irregular properties of BMR as observed in the forms of their respective distributions are the major causes of the observed variability in the polar field and the solar cycle. 
Moreover, our simulations show that the magnetic field-dependent time delay of BMR emergence produces more realistic features of the magnetic cycle, and this is consistent with observation as well.

\section*{Acknowledgements} 
The authors are thankful to the referee for carefully reviewing the manuscript and for providing insightful comments that helped to improve the quality of the manuscript to a large extent.
Authors acknowledge the Science and Engineering Research Board (SERB) for providing a financial support through the MATRIC program (file no MTR/2023/000670). 
The computational support and the resources provided by the PARAM SHIVAY Facility under the National Supercomputing Mission, the Government of India, at
IIT (BHU) Varanasi is gratefully acknowledged.

\section*{Data Availability}

 For all analyses, we have used data produced from the STABLE dynamo model and synthetic BMR code. Data from our models and the analysis codes can be shared upon a reasonable request.



\bibliographystyle{mnras}
\bibliography{paper} 








\bsp	
\label{lastpage}
\end{document}